# Investigating student understanding of quantum physics: Spontaneous models of conductivity

Michael C. Wittmann
Department of Physics, Department of Physics and Astronomy, 5709 Bennett Hall, University of Maine, Orono, ME 04469-5709

Richard N. Steinberg
Department of Physics, City College of New York, Department of Physics, CCNY, 138th St. at Convent Ave., New York, NY 10031

Edward F. Redish
Department of Physics, University of Maryland, College Park, MD 20742-4111

Students are taught several models of conductivity, both at the introductory and the advanced level. From early macroscopic models of current flow in circuits, through the discussion of microscopic particle descriptions of electrons flowing in an atomic lattice, to the development of microscopic non-localized band diagram descriptions in advanced physics courses, they need to be able to distinguish between commonly used, though sometimes contradictory, physical models. In investigations of student reasoning about models of conduction, we find that students often are unable to account for the existence of free electrons in a conductor and create models that lead to incorrect predictions and responses contradictory to expert descriptions of the physics. We have used these findings as a guide to creating curriculum materials that we show can be effective helping students to apply the different conduction models more effectively.

# I. Introduction

With the growing importance of electronic technology in our lives, it is clear that topics such as nanotechnology, photonics, superconductivity, and matter interferometry will be of increasing importance for engineers and materials scientists. Understanding these and related topics requires a good understanding of quantum physics. Yet many engineering schools are dropping modern physics requirements in order to shorten the time to graduation or to make room for computer classes.

Research in physics education at the introductory level indicates that students often leave physics service classes for engineers with less of an understanding of physics than we had expected or hoped.[1] A major problem is an "impedance mismatch" between student and instructor. What the student brings into the class affects how they interpret (or misinterpret) the information presented. When the instructor is unaware of the student's incoming state, instruction can be decidedly less effective.

It is possible that the traditional quantum mechanics course for engineers has fallen out of favor partly because it does not appear to provide enough value within a tightly constrained student curriculum. We hope that developing a course that fits well with the students' initial state and matches their motivational goals will prove more successful.

Until recently, there had been little physics education research in advanced topics such as quantum mechanics. In 1997, the authors and their collaborators began a research project to study the initial state and response to instruction of students in an upper-division one semester modern physics/quantum mechanics course for engineers. The project involved both education research and materials development.[2] The primary class considered is Physics 420 at the University of Maryland, a class that is usually taken in the junior or senior year mostly by electrical engineers (~85%). It is offered in both semesters and averages 12-25 students per semester. The students have studied and used differential equations extensively, but tend to be underprepared in matrix methods. In this paper, we report on our study of students' initial state and learning of an important topic for electrical engineers, conductivity.

The conduction of electrical current is a complex phenomenon. Physicists and engineers freely use a variety of different and sometimes contradictory models. One model treats conduction as a *macroscopic* phenomenon, describing it by macroscopic variables such as current and voltage. A second model provides a microscopic description of conduction using classical *free electrons* moving through a (nearly) fixed atomic lattice. A third provides a microscopic quantum description using *band structures* and de-localized electrons to describe conductive properties of the system. All these models are useful and meaningful. Students need to understand them and be able to use them where appropriate.

One way to explore student reasoning about the conducting properties of materials is to ask them about different circuit elements such as resistors, insulators, and wires. As students, particularly in electrical engineering, advance in their studies, they are asked to study and analyze these elements. We focus here on student reasoning about the material properties of conductors, insulators, and semiconductors, and how these properties lead to different conductive behavior. We describe a cyclical process of research, curriculum development, and evaluation of our work in order to create a curriculum best matched to our population of students. More information on this curriculum can be found on the Web.[2]

**Models of conductivity**

Models of conductivity are at times contradictory, yet students are supposed to know when and how to use the various models. We begin with a brief discussion of three models and their cognitive character.

In the *macroscopic model*, the system is treated as if it consists of macroscopic objects such as batteries, power sources, resistors, capacitors, transistors. This simplest, least structured model can be treated as if it had no underlying structure and uses macroscopic variables such as charge, current, and voltage. The objects are specified via measured parameters and functions such as resistance, capacitance, and constitutive relations (Ohm's law and the capacitance relation). Although charge is often described as being made up of electrons and ions, the properties of these electrons and ions play no essential role in the student's facility with the model.

The second model is more explicitly reductionist. In what we will refer to as the *Drude model*,[3] matter is treated as made up of ions and free classical electrons. When taught this model, students are typically presented with the representation of electron behavior shown in Figs. 1a and 1b.[4] Charge carriers are conceptualized as free particles moving with a velocity in a particular direction, as shown in Fig. 1a. With appropriate sign conventions, current is defined as the rate at which charge flows through some area. Next, or sometimes simultaneously, students are presented with the description of the random motion of electrons and how the electron drifts in the presence of an electric field (for example, described by the voltage of the system), as shown in Fig. 1b. In studying resistance and energy in simple circuits, an analogy is often made to mechanical systems such as water or marbles encountering physical obstacles. Charge carriers gain energy from a source and lose energy when they encounter obstacles. Throughout this stage of instruction, electrons are thought of as free particles without regard to the atomic lattice, except when the collisions depicted in Fig. 1b are described as collisions with the atoms. The simple assumption that there is an average time between randomizing collisions allows one to derive Ohm's law (although the numerical parameters that result do not arise in a natural way or appear consistent with simple classical assumptions.) In general, students should be able to use the Drude model to describe electron behavior, understand common representations, and reason with it. For example, if the temperature of a material is increased, vibrations in the atomic lattice shorten the average electron mean free path and create more resistance and lower the conductivity.

The third and most sophisticated model requires an understanding of band structures and bulk properties of matter. We will refer to this model as the *band structure model*. In this model, the quantum character of the system is an essential component. Individual conduction electrons are typically de-localized and the general properties of materials such as carrier concentration play a role. Electrons are depicted as particles moving within a band or between bands, as shown in Fig. 1c for a semiconductor. The movement of electrons between bands is a fundamental but subtle adaptation of the model of conductivity portrayed in Figs. 1a and 1b where electrons move in a purely classical way. Conduction occurs through a biasing on the population in an open band to states moving in one direction rather than the other. The motion is brought to a steady state by interactions with phonons (exchanging energy with the lattice of ions), a process that plays the role of ionic collisions in the Drude model. In general, students should be able to explain different types of band structures and how common representations of band structures are consistent with the existence of charge carriers in a system. Furthermore, they should be able to describe band structures of semiconductors and how temperature plays a role in carrier concentration in materials such as doped and undoped silicon.

These models relate to each other, as do thermodynamics, kinetic theory, and a full quantum statistical treatment. Just as students learning concepts in thermodynamics frequently confuse internal energy, heat flow, and temperature, we expect that a similar situation takes place when students learn about conductivity. When advanced engineering topics such as diodes and transistors are discussed in engineering classes, all three models are typically invoked.[5]

**Physics education research and conductivity**

Although the sequence of models presented in the previous section can lead to a robust and functional model of conductivity, there is great potential for confusion. Research in physics education has revealed that what students learn is often very different from what is presented to them.[1, 6] Even when instruction is complete and accurate, students often leave our physics classes with ideas that are in stark contrast to the way physicists think.[7]

For example, after completing instruction on dc circuits in introductory calculus-based physics, only about 15% of students are able to correctly rank the brightnesses of the bulbs in the circuit shown in Fig. 2.[8] Explanations given by students for the incorrect rankings to this and other simple circuits reveal fundamental problems with the way students think about conduction. Many say that "current is used up" in bulb A so that bulb C will be less bright. Others will use the words "voltage," "current," and "power" interchangeably or will view a battery as a constant current source regardless of the circuit. These types of difficulties are prominent after instruction regardless of the proficiency of the instructor, the background of the students,[9] or whether the students have completed a traditional circuits laboratory.

Further research has shown that students are often unable to build adequate links between the macroscopic and Drude models when trying to analyze specific phenomena in the physics of circuits.[10,11] University students were asked about a variety of simple RC circuits in situations where capacitors were being charged or discharged. Most were unable to define or describe the capacitance of an open switch. When asked to reason about transients in a circuit, students who had not received targeted instruction in microscopic models of current (for example, the Drude model) were more likely to misinterpret the meaning of memorized equations, emphasize the order of circuit elements inappropriately, or confuse current and voltage. Furthermore, students were often unable to give microscopic explanations for macroscopic behavior (for example, charging a capacitor). Students in a modified class[12] were far less likely to make these mistakes and were more likely to connect their understanding of charge with their understanding of circuits, current, and voltage. We note that the instructional situation is not yet fully understood and that other researchers have found contradicting results with high school students. For example, Gutwill et al.[13] found that targeted instruction with an emphasis on creating bridges between the microscopic and macroscopic models led to a decrease in student understanding of the physics in comparison to students who had learned the two models separately without explicit instruction in how they might complement each other.

Student difficulties with conductivity continue even in higher-level courses. For example, students studying the photoelectric effect in a sophomore level modern physics course demonstrated fundamental misunderstandings.[14] After instruction on the photoelectric effect, students were presented with a schematic diagram of the experiment, asked to draw a current-voltage graph, and account for their reasoning. Only about 25% of the students drew a correct graph and many of the incorrect answers revealed a weak conceptual model of conductivity. About one-third of the students drew a line through the origin. To them, Ohm's law governed all I-V behavior, even in the photoelectric effect experiment where there is an evacuated tube within the circuit. Another example of an incorrect response was the identification of the photons as the charge carriers between the two electrodes.

When the more sophisticated model portrayed in Fig. 1c is considered, there is even greater potential for confusion, especially given that many students enter this stage of instruction with the difficulties described above. One reason is that the model is fundamentally a semi-classical description. Band diagrams are predicated on the wave nature of the electrons, but the electrons are described as particles within the band diagrams. Furthermore, the shift in representation of electrons moving in just a spatial diagram to an energy-spatial diagram is potentially very confusing.[15]

## II. Student models of conductivity: focusing on single atoms

To understand how students in advanced physics classes understand the physics of current flow, we prepared a series of interviews, conceptual surveys, and examination questions. The particular type of interview we use is called a *demonstration interview*. It was used extensively by Piaget with physics questions at the primary and secondary school level[16] and extended to the college level by McDermott.[17] Students are shown a set of objects or an apparatus and asked questions that are designed to elicit their reasoning. In our case we are particularly interested in the physical models they use to make sense of the material. Because our goal is to understand student reasoning about the physics in detail, a variety of methods are required to gain a complete understanding of the typical students in our classroom. We begin with interviews because the interviewer is able to ask follow up questions to probe student reasoning more deeply than is possible with a typical written question. We follow with open-ended written questions that, although follow up is not possible, probe a broader segment of the class. Our survey results, not reported here, are consistent with those from interviews and examination questions. In the interviews, student volunteers usually rank at or near the top of their class, because students who are performing poorly are less likely to volunteer.

**Interview task**

The goal of the interviews was to probe student understanding of conductivity. Students were presented with tasks in which they had to make predictions and explain their reasoning in real contexts. In this way, the focus was on inferring student understanding from how they actually described physical systems. Thirteen students from two different Physics 420 classes were interviewed in detail on the subject of current flow. Nine were interviewed before any physics instruction on conductivity, while four were interviewed after all (unmodified) instruction in quantum mechanics, including several weeks of discussing quantum mechanical models of conductivity.

Students were asked to describe what would happen when a variety of materials were placed between two leads connected to a battery (see Fig. 3). Materials included steel wire, copper wire, aluminum, a rubber band, Styrofoam, pencil lead, and wood (all roughly of the same size). An example of a response would have been to state that in a metal, there are electrons in a conduction band, and these electrons are free to move about the material. When an outside voltage is placed on it, the electrons are biased to flow in a given direction. In such a situation, the electrons flow through a lattice of atoms, colliding with them, and reach a steady state average speed. Nearly all the students were also asked about the effect of repeating the experiment (with all of the elements) in an oven, where the temperature of the wire would be higher.

By comparing current flow in different materials (such as steel and copper, or copper and a rubber band), we were able to probe how students distinguished between the two materials and the different current flow characteristics in each. By asking about the experiment at different temperatures, we were able to probe how students applied their reasoning to make predictions of slightly different systems.

*Overview of student responses*

Students at all levels of instruction gave incorrect predictions and incomplete descriptions of the physics. We summarize the most common incorrect responses by focusing on student thinking about free electrons in a material. Rather than accept the existence of free electrons (as is done in the Drude model) and reason from that perspective, students often spontaneously made the connection to their model of individual atoms in order to provide a source of electrons. Of the thirteen students, seven (including post-instruction students) described variations on the idea that the "energy" (or "power" or "voltage") of the battery pulled the bound electrons off the atoms, allowing them to move through a wire. Movement of electrons then consisted of the electrons jumping back into bound states (in "holes" created by other electrons having been pulled off the atom), being pulled back out, and so on.

Broadly speaking, students did not think of the bulk properties of the system (often referring to questions about the origin of electrons in the system with "that's chemistry, isn't it?"). Instead, they built models of the situation by focusing on individual atoms.

The types of student responses to the interviews were the same, whether they came before or after instruction in Physics 420, though their frequency was different (as can be expected with differing levels of instruction and with such small student numbers in each). The types of responses given in the interviews, examinations, and surveys were also independent of student major, so that biologists, education students (specializing in math and physics), mechanical, electrical, and chemical engineers all gave the same, atom-based view of electron conduction. We find the consistency of the students' responses interesting because the generative, productive fashion of their thinking reveals the basic ideas that students use to make sense of what occurs in our classes, regardless of their previous instruction.

We next describe several explicit examples and the problematic predictions students made when using this model. We hope to illustrate through the extended discussion of student comments how their reasoning about the physics is at odds with what we are trying to teach them.

*Electron pull description: electrons pulled from individual atoms*

The most common description given by students of free electrons in a system is shown in Fig. 4. In this model, an applied electric field acts on individual atoms in such a way that electrons from the outer shell of the atom are "torn off."

Several students gave variations on this description. One student, David,[18] was asked, "Do all of the electrons inside the iron [of the wire] move [when the wire is attached to both leads]?" He replied, "For electrons, in order to flow through the wire, they have to leave the atom. You have to offer them enough energy in order to escape from the forces keeping them in the atmosphere of the atom, and then they move freely. They gain that energy, they come out from the structure of the atom, and then they move freely." When asked if there were any free electrons before the battery was attached, David replied, "I think no."

We refer to this response as the *electron pull description*, because students giving this description focused on individual atoms talked about pulling electrons away. Sarah noted, "it takes certain energies to tear [electrons] away," while Thomas stated, "just the [electrons] on the utmost outer shell would move, and they'd get pulled off the atom." We have also observed this explanation with students trying to explain conduction in a classroom situation.[19]

*Atomic jump description: Electrons jumping from bound state to bound state*

Students who gave the electron pull description of free electrons commonly used two different descriptions of electron flow. In each, electrons pulled from atoms were re-absorbed into other atoms. In the first, atoms that had lost electrons previously had a "hole" (in their shell) to be filled. Another freed electron filled the hole. Electrons were then pulled from the atom to create another hole and create electron flow. In the second, a free electron enters into an atom that has not previously lost an electron. The additional electron's presence creates an imbalance, which forces an electron out of the atom.

David illustrates how students used the first description, with electrons jumping into previously vacated holes, only to leave again. He stated (see Fig. 5 with annotations as to which electrons he pointed to while speaking): "You have this process: the electrons are moving in this direction, they get out from one atom, the electron gets out from here [atom A], let's say, the electron is moving … This electron takes the place of this electron here [atom B], this one takes the place of this one [atom C], and then this one moves to this place [not included in diagram], then it moves to the next place, and it comes again out of the atom and it moves to the next." Interviewer: "So it becomes part of the atom for some amount of time, and then it leaves again?" David: "Yes. Some professors in electrical engineering, they like to describe this thing like you have holes that are left empty when the electron comes out, and these electrons come out and move and fill the holes, and then they move, and if there is a hole in their path, they fill the hole and then they come out and another electron comes in that hole."

Note that David has tried to interpret what he has learned in previous classes, stating, "you have holes that are left empty" and describing a method whereby holes flow in the opposite of the electrons in a circuit. As with many other students in the interviews, David has found a way to create (at least partial) consistency with the general description often used in his previous classes (the flow of holes opposite that of electrons) and his own model of the physics of conductivity.

Peter illustrates the second point. He used a diagram nearly identical to that drawn by David, as did several of the students. When asked, "what happens inside the wire" when current flows, he stated, "There's [sic] obviously electrons in the wire. And what will happen is that basically, there will be a chain reaction of electrons going through, … this one electron will kind of be projected onto the next atom, or whatever, it becomes unstable. And then, kind of a bouncing effect all the way through the circuit. And that's pretty much what the current flow is. … You pretty much have an electron being absorbed by a material, and then another electron emitted, so you kind of have a chain reaction going through the circuit."

As with David, the electrons are emitted (for example, pulled from an atom) and move for some distance before being absorbed into an atom. We refer to both of these descriptions (though they have obvious differences) as the *atomic jump* description of electron flow.

**Student predictions of properties of conductivity**

We emphasize that the incorrect student descriptions of conduction correspond to incorrect predictions about the effect of changes to the interview task. Students using the electron pull and atomic jump descriptions were asked to make predictions of the conductivity of different substances at different temperatures. They were also asked, both directly and indirectly, about the consistency of their answers and other models they used in their physics and engineering courses. We summarize their responses below.

*High temperature leads to higher conductivity in wires*

Many of the students were asked to describe the effect of placing the demonstration apparatus in an oven. An appropriate use of the Drude model would have led students to describe vibrations in the atomic lattice leading to shorter mean free path lengths, which would lower the conductivity of the wire. Instead, students who used the electron pull explanation often stated that heat added energy to the electrons, so that the "outer electrons" were easier to remove from the atoms. Thus, their prediction was that wires at higher temperature would have higher current flow than wires at colder temperatures. This prediction is exactly the opposite of what the Drude model predicts (and what happens).

David clearly stated the manner in which the electron pull explanation played a role, saying that "I know that heat is a form of energy, and if we assume that this heat, this energy that comes from the heat weakens the force that keeps the electron in the orbit of the atom, then they come out of the atom and they move more easily."

Peter also illustrates how students using the electron pull and atomic jump descriptions can come to the incorrect conclusion about the role of heat in creating more electron flow in a system. He stated, "When you raise the temperature, you get more atomic movement, and so that would probably aid in the transfer of electrons, so I'd say that the resistance would go down … I guess the atoms in a more excited state would promote the electronic transfer." Note that Peter clearly talks about atomic movement within the system, but does not apply it to the atomic lattice. Instead, he focuses on the electronic transfer, as if the atom were shaking electrons loose. It should be noted that he correctly described the effect of length and cross sectional area in affecting the resistance of a wire.

As with other interview excerpts, these students clearly state the common view held by many other students. Both predict that a higher temperature leads to lower resistance in the wire, because there would be more current flow due to increased numbers of electrons. Thus, the electron pull and atomic jump descriptions lead them to incorrect predictions of the physics.

*Breakdown in Ohm's law*

Many students were not able to interpret Ohm's Law correctly, nor to infer its implications in a microscopic setting. These students stated that there was a cut-off or threshold voltage required before current flowed in a wire. In other situations, though, these same students had no problems correctly applying Ohm's Law. Students seemed to reason independently when using the macroscopic model (such as describing current or resistance) and a microscopic model (such as average electron speed).

Two students, Thomas and David, clearly indicate that Ohm's Law does not guide their reasoning. Thomas stated, "Even though the steel wire is not a resistor, it still has its own internal impedance of how much energy it takes to remove the electron and then get it to move around." When asked if there would be no current flow at some small voltage, he says it's possible, explaining, "because there wouldn't be enough energy to remove the electron from its orbit."

Similarly, David was asked in the context of the wire in the circuit if there is "a voltage where I will not get any current flowing." He responded, "Yes, there is. If the voltage applied to this conductor is not enough to take the electrons out of the atoms, you do not have any electrons flowing in the wire, so this voltage has to have at least a minimum value for this voltage in order to see electrons flowing through the wire."

Both Thomas and David seem to be describing a step in the I-V curve, where $I = 0$ while $V \ne 0$ at low V. This topic was not explicitly part of the interviews, and few student interviews arrived at this point, but Thomas and David gave the common incorrect predictions for students who invoked Ohm's Law.

*Inability to reason about semiconductors*

A striking result of the interviews was student inability to discuss semiconductors. Eleven of the thirteen interviewed students had completed electrical engineering classes in which semiconductors were discussed and used, but only two were able to describe the band diagram representation of semiconductors in any detail. Several students used the electron-jump model described above. For example, Thomas explicitly used the electron pull description when he described the difference between doped and undoped semiconductors. He stated, "I think the doped ones are better conductors because I think it takes a lot of energy to remove the silicon electrons, but if you add electrons from a different metal, like aluminum, which require less energy to be removed, then you'd get more current using less energy." Here, the electron pull explanation allowed him to make an incomplete but correct prediction about the physics.

We note that the eleven electrical engineering students stated that they had discussed semiconductors in previous classes. A reasonable explanation for their responses in these interviews would be that they were unfamiliar with the physics of semiconductors and had so far learned to deal only with specific examples in their engineering classes.

*Spontaneous student models of conductivity*

Most students did not enter the interview with a coherent model of free electrons existing in a material. These students created a model of conduction that could account for both their existence and their motion during the interview. It was often noticeable that students were inventing responses to situations they appeared never to have considered before. To document the contradictions in their reasoning as it developed through the course of the interview is not the focus of this paper, but nearly all students used statements such as "I never thought of that before," or "good question, let me think about that." Thomas at one point stated, "I'm sorry, I didn't mean – I was wrong when I said earlier that all of [the electrons] move." Sarah first described atomic lattice vibrations in a heated wire impeding electron flow, but then changed her response to say that the energy, when transferred to electrons, helped the electron flow. This change in her response was brought about specifically by her development of the electron pull description during the interview in response to a request to account for her original answer. The electron pull description was created slowly, and she appeared to have never explicitly stated it before. Yet, she and many other students created the description, and in her case this thought had the effect of reversing a previously correct although incomplete prediction about the physics.

Strikingly, most of the students' spontaneous descriptions were similar. This similarity suggests that the reasoning that students use to arrive at their (often incorrect) answers is common. These findings are similar to results from investigations in other areas of physics.[20,21]

## III. Curriculum Development

Research-based curriculum design has been used to great effect for introductory course materials, for example, wave physics,[22,23,24] electrical circuits,[25] optics,[26] and Newtonian dynamics,[27] and for advanced course materials, for example, relativity,[28] the photoelectric effect,[14] and the wave properties of matter.[29] Typically, students work through materials designed to raise issues shown to be difficult in the interviews.

The materials described in this paper were created as part of the development of the *New Model Course in Applied Quantum Physics*.[2] The materials include small group learning environment worksheets (inspired by the University of Washington *Tutorials in Introductory Physics*[30]), associated conceptual homework problems, applied homework problems (in which students use concepts to analyze more complicated situations),[31,32] daily essay assignments (which are used in the mode of *Just-in-Time-Teaching*[33]), and conceptual examination questions. In our Tutorials, we use specially designed questions and relevant software tools to help students observe, discuss, and build an appropriate understanding of the physics. The conceptual homework problems are designed to help students practice individually what they learned in groups in the classroom. The applied homework is designed to show them the relevance of this material by emphasizing how one can understand familiar materials or objects using ideas from the Tutorial. In our essay assignments, we made a concerted effort to make further connections between classroom physics and the real world, while also emphasizing the different models used in understanding quantum physics. The effectiveness of this approach was then tested by the use of examination questions.

### Design of curriculum materials

The unit on models of conductivity includes three Tutorials, three conceptual homework assignments, one applied homework assignment, several essay questions, and an examination question designed to evaluate student learning. We give an outline of the final version of the curriculum materials as they were redesigned based on the student interviews described above and also based on examination results described below.

Students begin by building a model of band diagrams and discussing polarization of a metal using a simple one-dimensional model of finite square and then Coulombic wells. During this Tutorial, students use elements of *Visual Quantum Mechanics*[34] and the *CUPS* utilities[35] to assist students in visualizing the descriptions they give. Students use concepts of bands of energy levels in small, one-dimensional atomic lattices to describe band diagrams for conductors, semiconductors, and insulators and the simple descriptions of the charge carriers in each case.

Only after they have described the origin of free electrons, do they return to describing the motion of free electrons. At this point, they change physical models and describe electrons as particles flowing through an atomic lattice. Once they have carried out these activities, they discuss which model applies and which is more appropriate to describe a given situation. The Tutorials help students evaluate when the Drude model appropriate and when the band structure model is appropriate. This skill is assumed in many other curricula, but here we have made it an explicit part of instruction. For each Tutorial, there is an associated conceptual homework assignment. In addition, at the end of this discussion, students were given homework in which they used their models of conductivity to describe the physics of a pn-junction. Throughout this process, they answered essay questions that built on discussions from the Tutorial and lectures. Evaluation, described below, was carried out by the use of an examination question.

**Preliminary Evaluation of Student Performance: Comparison of Three Classes**

To evaluate the effectiveness of these materials in helping students understand representations and descriptions of the physics of conduction, we asked an identical examination question in three different Physics 420 courses (see Fig. 6). The question was designed to investigate whether students could reason using both band diagrams and the Drude model. When students answered part (a) on band diagrams, we looked both at their multiple-choice responses and at their reasoning. When these two conflicted (for example, giving "D" for the conductor, but describing a semi-filled conduction band), we categorized student answers in terms of their other responses to the problem. Usually, more emphasis was placed on their reasoning than their multiple-choice response, because a response alone does not indicate enough about how they arrived at their answer.

The three courses came in consecutive semesters. The "traditional" course received only traditional instruction, with no specially designed materials. Lecture topics included Fermi energies and models of heat and current conduction. The instructor in this course focused on conductivity for the last 3 weeks of the semester, with three one-hour lectures each week. The two other classes had modified instruction (using elements of our *New Model Course* which were in development at the time). The first class (modified 1) used an early version of the Tutorials with no essays or applied homework problems. In these Tutorials, the more common order of instruction was used, where students described the flow of free electrons through a wire before describing band diagrams of common substances. The second (modified 2) used the Tutorial sequence described at the end of the previous section (including tutorials, essays, and applied homework). In both classes, Tutorials were taught in the place of one lecture (of three) a week. The total amount of time spent on conductivity was the same in both the modified and the traditional classes.[36]

Figure 7 shows results from the three classes. Each bar indicates the percentage of students who correctly answered only the band diagram part of the question (part a), only the Drude model part of the question (part b), or both. For example, in the traditional class, 58% answered only the band diagram part correctly, 33% answered only the Drude model part correctly, and 25% answered both correctly. Thus, slightly less than half the students giving a correct answer on the band diagram section were also able to answer the Drude section correctly. The performance of students using the original tutorial materials (modified 1) was slightly worse than that of students in the traditional class. Although the total number of students using either model was slightly greater than in the traditional class (71% versus 64%, respectively), fewer students were able to describe both the Drude and the band diagram models correctly (8% versus 25%, respectively).

Because of the weak performance of students using the original tutorial materials, we modified them to better match our findings from the interviews. Students were given the opportunity to develop their reasoning about the source of free electrons first. Also, essay questions throughout the semester (including during the instruction on conductivity) asked about situations in which multiple models might be required to describe a single situation. Only after these changes were made, were the modified 2 students more successful in answering the question. Note that we measure success both in terms of how many students answered correctly and how many were able to correctly apply two different models in the question. We also emphasize that these results are preliminary, and further study is required to test whether the materials are effective.

## Discussion

The process of physics education research, specifically the use of research into student reasoning about a topic to develop and revise targeted curriculum materials, can lead to improved student learning of advanced physics topics. In addition, this research helps us come to a better awareness of the nature of student thinking and how students make sense of the material that they are learning. By recognizing the nature of student reasoning about the source of free electrons in a metal, we have developed appropriate materials to address their needs. Giving students an opportunity to develop tools for understanding the physics has had a measurable effect on student learning. These materials form only a small part of an entire course that we have developed, but are representative of how research based curriculum development can create a more effective learning environment for students.

## Acknowledgements

This work took place within the Physics Education Research Group at the University of Maryland, and we are indebted to all members of the group for discussions and assistance in our work. In particular, we would like to thank Lei Bao for helping with early investigations into this topic, Rebecca Lippmann for helping with the curriculum development, and the instructor in the traditional course for allowing us to observe and investigate his class. We also thank the anonymous reviewer for valuable comments and suggestions. Finally, we wish to acknowledge grants from the National Science Foundation (DUE-965-2877) and the Department of Education Fund for the Improvement of Post-Secondary Education (P116B970186 and P116B000300), which provided financial support.

# Figures
**Figure 1. Typical presentation of models of conductivity in an introductory physics text.**

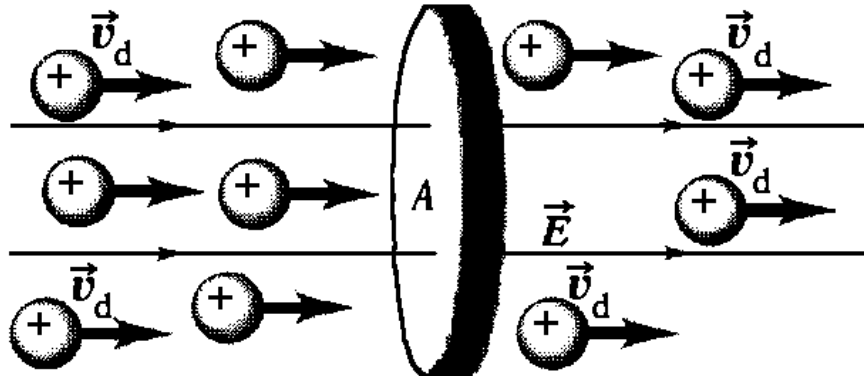

(a) Current presented as the flow of charge through some area.

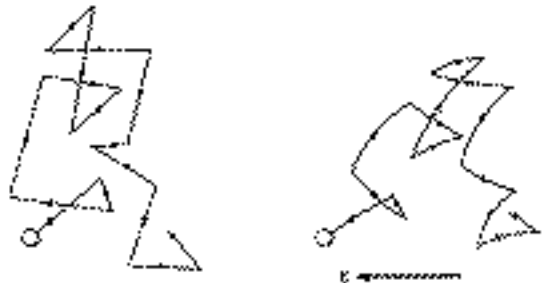

(b) Random motion and drift of electrons.

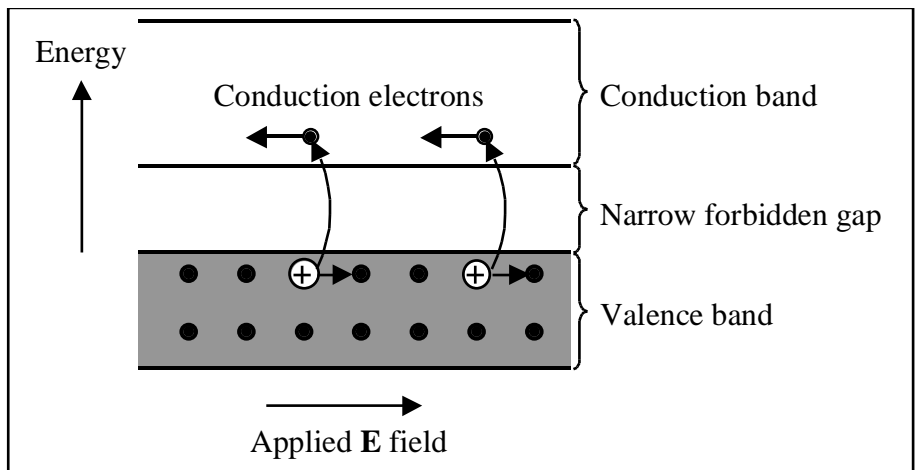

(c) Electron motion within a band diagram description.

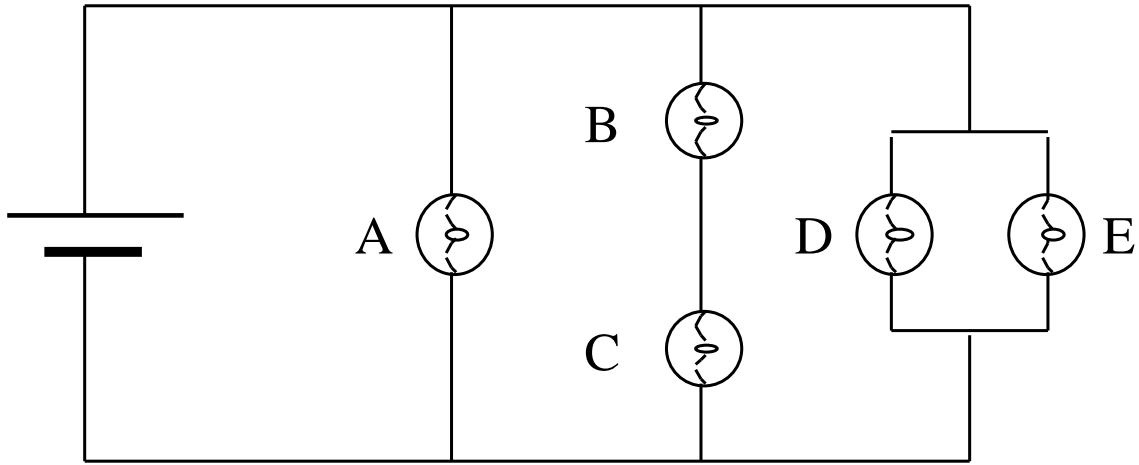

**Figure 2.** Circuit of 5 identical bulbs connected to a battery, presented to students after instruction in dc circuits. Fewer than 15% of the students can rate the brightnesses correctly.

Present students with:

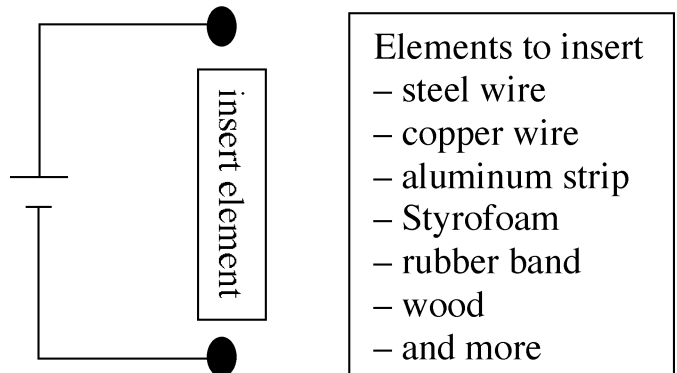

Elements to insert
– steel wire
– copper wire
– aluminum strip
– Styrofoam
– rubber band
– wood
– and more

**Figure 3:** Interview task. Students were provided with a battery and a variety of materials that could be placed between two leads and were asked to describe what would happen in each case.

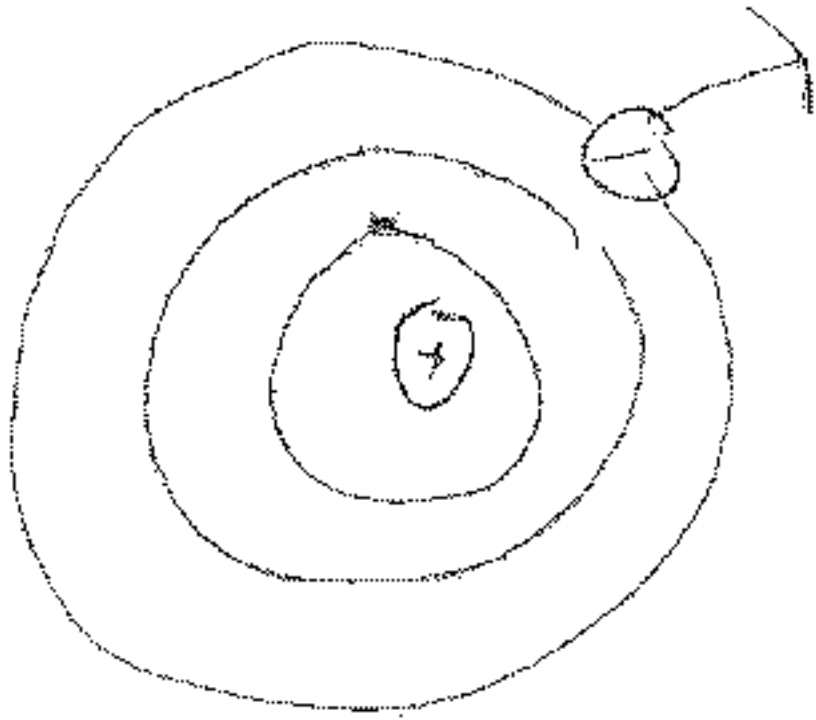

**Figure 4: "Electron pull description"** as drawn by a student. Students describe electrons in the outer shell of an atom getting pulled off by the "energy" given by the battery in a circuit: "Just the [electrons] on the utmost outer shell would move, and they'd get pulled off the atom."

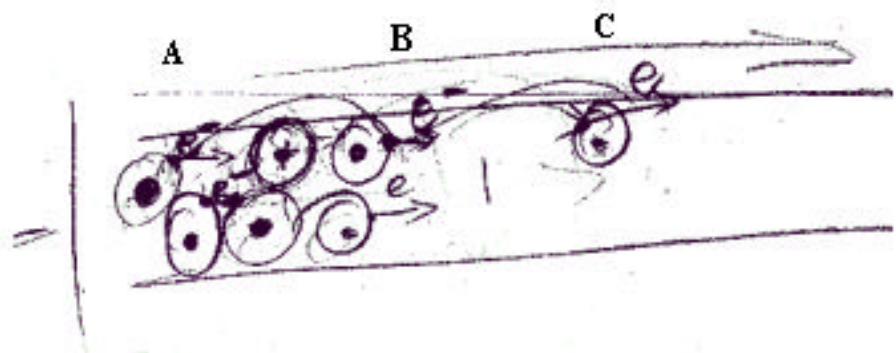

**Figure 5: "Atomic jump description"** of electron (and hole) flow in a wire, as drawn by a student. Students describe electrons jumping from bound state to bound state.

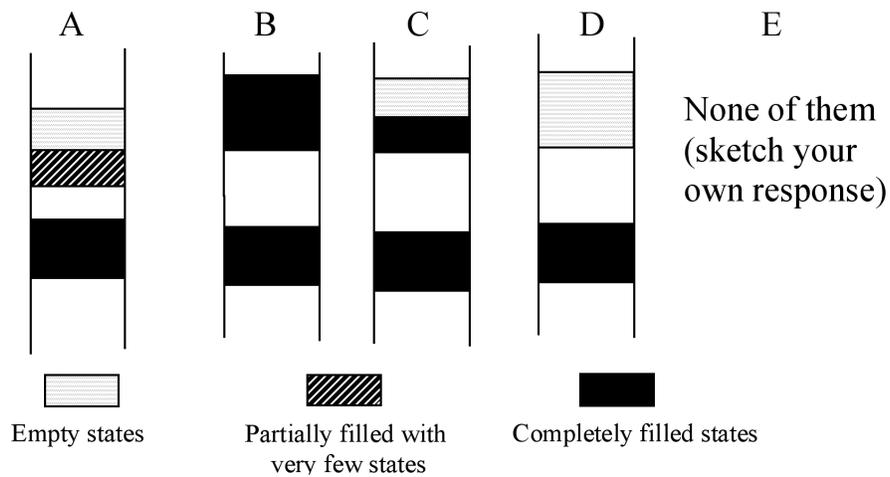

a. Consider the following band diagrams. For the each of the materials listed below the diagram, select the diagram that best represents the band structure of the material. For each material, **explain how you arrived at your answer.**
   i. Conductor      ii. Insulator      iii. Semiconductor

b. A potential difference is placed across a resistor. Describe what, if anything, happens to the individual electrons in the material. Include a sketch in your explanation.

**Figure 6: Final examination question testing student understanding of band diagrams and the Drude model. Student responses to the multiple-choice section were interpreted based on their written explanations.**

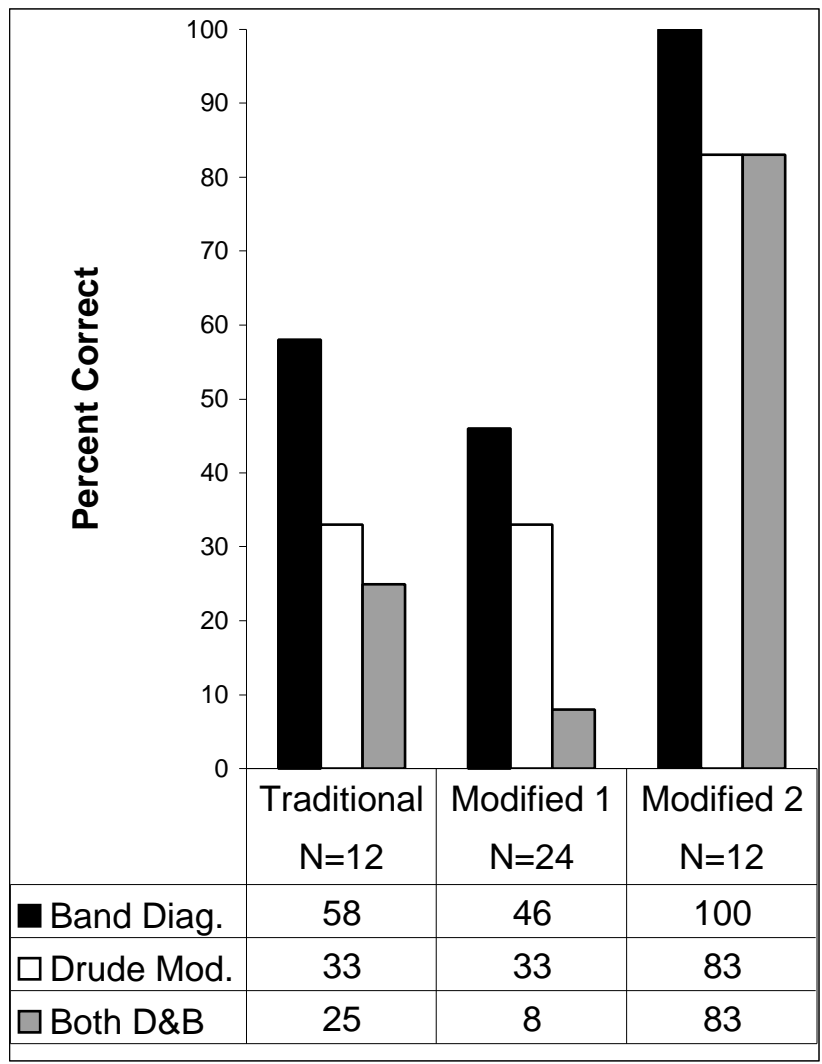

**Figure 7: Student performance on examination question shown in Fig. 6. Both modified instruction classes used Tutorials, while Modified 2 used additional materials. All three classes spent roughly the same amount of class time on the topic of conductivity.**

[1] Lillian C. McDermott and Edward F. Redish, "Resource letter PER-1: Physics education research," Amer. J. Phys. **67**, 755-767 (1999).

[2] Edward F. Redish, Richard N. Steinberg, and Michael C. Wittmann, "A new model course in applied quantum physics," http://www.physics.umd.edu/perg/qm/qmcourse/welcome.htm. These materials are freely available.

[3] E. Whittaker, *A History of the Theories of Aether and Electricity, Vol. 1: The Classical Theories* (T. Nelson and Sons, Ltd., London, 1951).

[4] Figure 1a is from Hugh D. Young and Roger A. Freedman, *University Physics* (Addison Wesley, New York, 2000), 10th ed. Figures 1b and 1c are from Raymond A. Serway, *Physics for Scientists and Engineers with Modern Physics* (Saunders College Publishing, Philadelphia, 1996), 4th ed.

[5] Ben G. Streetman and Sanjay Banerjee, *Solid State Electronic Devices* (Prentice-Hall, Inc., Englewood Cliffs, NJ, 2000), 5th ed.

[6] Edward F. Redish and Richard N. Steinberg, "Teaching physics: Figuring out what works," Physics Today **52** (1), 24-30 (1999).

[7] Lillian C. McDermott, "Millikan Lecture 1990: What we teach and what is learned – Closing the gap," Amer. J. Phys. **59**, 301-315 (1991).

[8] Lillian C. McDermott and Peter S. Shaffer, "Research as a guide for curriculum development: An example from introductory electricity. Part I: Investigation of student understanding," Amer. J. Phys. **61**, 994-1003 (1992).

[9] Even students at Harvard encounter similar difficulties after studying introductory circuits. See Eric Mazur, "Understanding or memorization: Are we teaching the right thing?," in *Conference on the Introductory Physics Course*, Jack Wilson, ed. (Wiley, New York, 1997), p. 113.

[10] Beth Ann Thacker, Uri Ganiel, and Donald Boys, "Macroscopic phenomena and microscopic processes: Student understanding of transients in direct current electric circuits," Amer. J. Phys. 67 (Physics Education Research Supplement), S25-S31 (1999).

[11] Bat-Sheva Eylon and Uri Ganiel, "Macro-micro relationships: the missing link between electrostatics and electrodynamics in student reasoning," Int. J. Sci. Educ. 12 (1), 79-94 (1990).

[12] These students used Ruth Chabay and Bruce Sherwood, *Electric and Magnetic Interactions* (Wiley, New York, 1995).

[13] Joshua P. Gutwill, John R. Frederiksen, and Barbara Y. White, "Making their own connections: Students' understanding of multiple models in basic electricity," Cognition and Instruction **17** (3), 249-282 (1999).

[14] Richard N. Steinberg, Graham E. Oberem, and Lillian C. McDermott, "Development of a computer-based tutorial on the photoelectric effect," Amer. J. Phys. **64**, 1370-1379 (1996).

[15] For additional information on this topic, see Edward F. Redish, Lei Bao, and Pratibha Jolly, "Student difficulties with energy in quantum mechanics," presented at the American Association of Physics Teachers Winter Meeting, Phoenix, AZ, 1997 (unpublished, available online at http://www.physics.umd.edu/rgroups/ripe/perg/quantum/aapt97qe.htm).

[16] See, for example, Jean Piaget and Barbel Inhelder, *The Child's Conception of Space* (W. W. Norton, New York, 1967), and Barbel Inhelder and Jean Piaget, *The Growth of Logical Thinking*

*From Childhood to Adolescence; An Essay on the Construction of Formal Operational Structures* (Basic Books, New York, 1958).

[17] See David E. Trowbridge and Lillian C. McDermott, "Investigations of student understanding of the concept of velocity in one dimension," Amer. J. Phys. **48**, 1020-1028 (1980), and David E. Trowbridge and Lillian C. McDermott, "Investigations of students' understanding of the concept of acceleration in one dimension," Amer. J. Phys. **49**, 242-253 (1981).

[18] Note that we use alias names throughout this paper. Furthermore, all quotes are from students who had not yet received instruction on conductivity in Physics 420, but two of the four students who had completed instruction gave answers consistent with those described here. All quoted students were in the top quarter of their class.

[19] A transcript of this discussion can be found in the example of student-student interactions during Tutorial instruction, shown in the teacher's guide section of Ref. 2.

[20] David Hammer, "Student resources for learning introductory physics," Amer. J. Phys. **67** (Physics Education Research Supplement), S45-S50 (2000).

[21] Michael C. Wittmann, "The object coordination class applied to wavepulses: Analysing student reasoning in wave physics," Int. J. Sci. Educ., in press (2001).

[22] Michael C. Wittmann, Richard N. Steinberg, and Edward F. Redish, "Making sense of students making sense of mechanical waves," The Physics Teacher **37**, 15-21 (1999).

[23] Michael C. Wittmann, Richard N. Steinberg, and Edward F. Redish, "Understanding and affecting student reasoning about the physics of sound," in preparation (2001).

[24] Diane J. Grayson, "Using education research to develop waves courseware," Computer in Physics **10**, 30-37 (1996).

[25] Lillian C. McDermott and Peter S. Shaffer, "Research as a guide for curriculum development: An example from introductory electricity. Part II: Design of an instructional strategy," Amer. J. Phys. **61**, 1003-1013 (1992).

[26] Karen Wosilait, Paula R. L. Heron, and Lillian C. McDermott, "Development and assessment of a research-based tutorial on light and shadow," Amer. J. Phys. **66**, 906-913 (1998).

[27] Lillian C. McDermott, Peter S. Shaffer, and Mark D. Somers, "Research as a guide for teaching introductory mechanics: An illustration in the context of the Atwood's machine," Amer. J. Phys. **62**, 46-55 (1994).

[28] Rachel Scherr, Stamatis Vokos, and Peter S. Shaffer, "Student understanding of relativity," Amer. J. Phys. **69** (Physics Education Research Supplement) S24-S35 (2001).

[29] Bradley S. Ambrose, Peter S. Shaffer, Richard N. Steinberg, and Lillian C. McDermott, "An investigation of student understanding of single-slit diffraction and double-slit interference," Amer. J. Phys. **67**, 146 (1999).

[30] Lillian C. McDermott, Peter S. Shaffer, and the Physics Education Research Group at the University of Washington, *Tutorials in Introductory Physics* (Prentice Hall, New York, 1998).

[31] Zuyuan Wang, Edward F. Redish, and Seth Rosenberg, "Quantum physics for engineers and applied physicists: Applied homework assignments," presented at the Proceedings of the ICPEC, Chungbuk, Korea, August 2001 (unpublished).

[32] The applied homework problems were inspired by work on introductory physics alternative homework assignments: Edward F. Redish and the University of Maryland Physics Education

Research Group, "Alternative Homework Assignments," available at http://www.physics.umd.edu/perg/abp/aha/.

[33] Gregor M. Novak, Evelyn T. Patterson, Andrew D. Gavrin, and Wolfgang Christian, *Just-in-Time-Teaching: Blending Active Learning with Web Technology* (Prentice Hall, Upper Saddle River, NJ, 1999).

[34] Dean Zollman and Kansas State University Physics Education Research Group, "Visual Quantum Mechanics," http://www.phys.ksu.edu/perg/vqm/

[35] John Hiller, Ian Johnston, and Daniel Styer, "Quantum Mechanics Simulations," in *Consortium for Upper-Level Physics Software (CUPS)*, Maria Dworzecka, Robert Ehrlich and William MacDonald, eds. (John Wiley & Sons Inc., New York, 1994).

[36] Note that the difference in class sizes does not reflect drop rates, but that the instructor in the "modified 1" class was well known and many students sought his class out. Drop rates in all classes were roughly equal at 20% from original enrollment.